\documentclass[
    ,final            % use final for the camera ready runs
  ]
  {aipproc}

\layoutstyle{8x11double}

\newcommand{\ie}{i.e.\ }
\newcommand{\GeV}{\,\mathrm{GeV}}
\newcommand{\fb}{\,\mathrm{fb}}

\newcommand{\LPTHE}{LPTHE;
    UPMC Univ.~Paris 6;
    Univ.~Paris Diderot (Paris 7);
    CNRS UMR 7589;
    Paris, France}
\newcommand{\UCL}{Department of Physics \& Astronomy, University College
  London.}

\begin{document}

\title{Jet substructure as a new Higgs search channel at the LHC%
\thanks{Talk presented by GPS at SUSY08, Seoul, Korea, June 2008.}
}


\author{Jonathan~M.~Butterworth}{address={\UCL}}
\author{Adam~R.~Davison}{address={\UCL}}
\author{Mathieu Rubin}{
  address={\LPTHE}
}
\author{\underline{Gavin P.~Salam}}{
  address={\LPTHE}
}

% \author{<author2>}{
%   address={<common address for author2 and author3>}
% }
% 
% \author{<author3>}{
%   address={<common address for author2 and author3>}
%   ,altaddress={<author1 address>} % additional visiting address
% }

\begin{abstract}
  These proceedings discuss a possible new search strategy for a light Higgs
  boson at the LHC, in high-$p_t$ WH and ZH production where the Higgs
  boson
  decays to a single collimated $b\bar b$ jet. Material is included
  that is complementary to what was shown in the original article,
  arXiv:0802.2470.
\end{abstract}

\maketitle

%%%%%%%%%%%%%%%%%%%%%%%%%%%%%%%%%%%%%%%%%%%%
%% MAINMATTER
%%%%%%%%%%%%%%%%%%%%%%%%%%%%%%%%%%%%%%%%%%%%

%\section{Introduction}

The search for the Higgs boson is one of the main priorities for the
Tevatron and the LHC. Current electroweak precision fits~\cite{Grunewald:2007pm}
suggest that the Higgs boson may be light, \ie with a mass $m_H\simeq
120\GeV$. This region of mass is one of the most difficult in which to
discover the Higgs boson, in part because its main decay channel, to
$b\bar b$, is swamped by large QCD backgrounds. 
Accordingly most search strategies rely either on looking for rare but
characteristic decay channels such as $H \to \gamma \gamma$, or
alternatively for $H\to b\bar b$ decays in production channels with an
associated, leptonically-decaying $W$ or $Z$ boson, which provides an
electroweak ``tag'' that is rare in backgrounds.

\begin{figure}
  \includegraphics[width=0.8\linewidth]{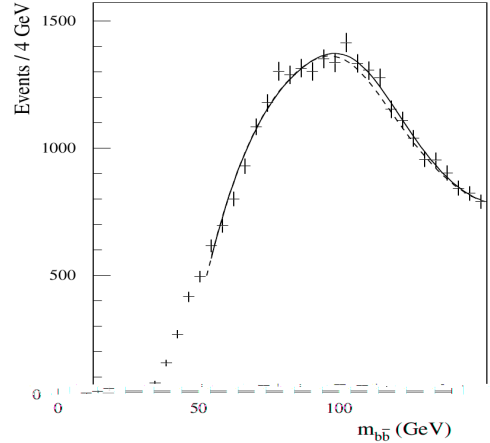}
  \caption{Background (dashed line) plus Higgs signal (solid line) for
    a Higgs boson with $m_H=100\GeV$ in the $pp \to WH$, $W\to
    \ell^{\pm}\nu$ and $H\to b\bar b$ channel, as found in the ATLAS
    TDR study \cite{atlasphystdr} for $30\fb^{-1}$.}
  \label{fig:atlasphystdr}
\end{figure}

While this second approach seems promising at the
Tevatron~\cite{Sanders:2008is}, studies from a few years
ago~\cite{atlasphystdr} suggested that it would be very challenging
at the LHC. The difficulty is clearly visible in
fig.~\ref{fig:atlasphystdr}, which shows simulated background (dashed
line) and background+signal (solid line) distributions.
Not only is the ratio of $S/\sqrt{B}$ rather low, but the signal
is a tiny addition to a background with strong kinematical structure
near the generated Higgs mass, an artefact due to
transverse momentum cuts in the analysis and a significant $t\bar t$
background.
Fig.~\ref{fig:atlasphystdr} implies a need for exquisite control of the
background shape if the Higgs boson is to be identified here.

%% The kinematical structure arises for two reasons: cuts on the
%% $p_t$ of the $b$ and $\bar b$ jets introduce a large mass-scale into
%% the problem, and one of the largest backgrounds, semi-leptonic $t\bar
%% t$ events, leads to events where the $b\bar b$ invariant mass is
%% naturally close to the Higgs-boson mass.

Recently we 
% it was  
suggested~\cite{Butterworth:2008iy} that the WH and
ZH channels might be recovered as potential discovery channels by
restricting one's attention to the $\sim 5\%$ of events in which the
vector and Higgs bosons each have a large transverse momentum, $p_{tV}
\simeq p_{tH}> 200\GeV$ and are back-to-back. As we shall discuss in
more detail later, this is advantageous (despite the large
reduction in signal cross-section) because it will greatly increase
the ratio of signal to background, and largely eliminate the problem
of the non-trivial ``shape'' of the background.

To investigate the potential of such a high-$p_t$ Higgs search,
ref.~\cite{Butterworth:2008iy} carried out a hadron-level analysis, which can be
factored into a leptonic side and a hadronic side. The leptonic side
is straightforward: electrons and muons are considered to be
identified if they have $p_t > 30\GeV$ and $\eta < 2.5$, missing
energy is considered if $/\!\!\!\!\!\! E_T > 30 \GeV$, and one looks
for events consistent with $Z\to \ell^+\ell^-$, $Z \to \nu\nu$ or $W
\to \ell^{\pm} \nu$ and places a cut on the total $p_t$ of the
vector boson, $p_{tV} > p_{t,min}$.

The hadronic side requires more sophistication if one is to maximise
signal to background ratios and obtain a good mass resolution. The
high-$p_t$ Higgs decays to a $b\bar b$ pair, which leads to a single
broad jet. Until recently, the state-of-the-art
approach~\cite{Seymour:1993mx,Butterworth:2002tt,Butterworth:2007ke}
for identifying such decays exploited the hierarchical nature of the
$k_t$ algorithm~\cite{Catani:1993hr,Ellis:1993tq}. This is effective
in rejecting backgrounds but suffers from poor mass resolution, while
split--merge-based cone algorithms often give better mass resolution
but with poor background rejection.

A powerful mix of the two approaches can be constructed~\cite{Butterworth:2008iy}
using the Cambridge/Aachen
algorithm~\cite{Dokshitzer:1997in,Wobisch:1998wt}, which successively
recombines the closest pair of particles (or pseudojets) in the event
until all are separated by a rapidity-azimuth distance of at least
$R$.
After having clustered the event and identified a hard massive jet,
one undoes one step of its clustering. This breaks the jet into two
subjets: if the heavier subjet is significantly lighter than the
original jet and the sharing of momentum between the two subjets is
not too asymmetric\footnote{``Significantly lighter'' and ``not too
  asymmetric'' involve cuts that can be chosen based on considerations
  of leading order QCD emission. The specific values of the cuts
  that were used are described in~\cite{Butterworth:2008iy}.} then one works with
the hypothesis that the two subjets correspond to the $b$ and $\bar b$
from the Higgs decay.
Otherwise one discards the lighter subjet and repeats the unclustering
procedure on the heavier subjet. 

\begin{figure}
  \includegraphics[width=\linewidth]{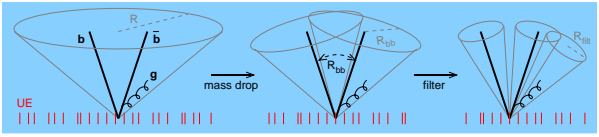}
  \caption{Mass drop and filtering procedure.}
  \label{fig:MDF}
\end{figure}

Once one has a Higgs candidate, one verifies that both subjets have a
$b$-tag. This together with the symmetry cut helps eliminate much of
the background. The mass resolution, however, is not very good at this
stage: by triggering on the mass drop, one has an effective jet radius
($R_{b\bar b}$) that corresponds closely to the ideal radius for
capturing all radiation from the decaying Higgs boson; but one also
captures much underlying event (UE), which degrades the mass
resolution. The next step is therefore to further undo the clustering
to an effective radius of $\min(0.3,0.5R_{b\bar b})$ and take the 3
hardest resulting subjets: the $b$, $\bar b$ and the hardest emitted
gluon. This eliminates much of the UE, while keeping most of the hard
perturbative radiation from the Higgs decay.

\begin{figure}
  \rotatebox{90}{\qquad\quad \small d$N/$d$m_{\rm H}$ [arbitrary units]}
  \includegraphics[width=0.45\linewidth]{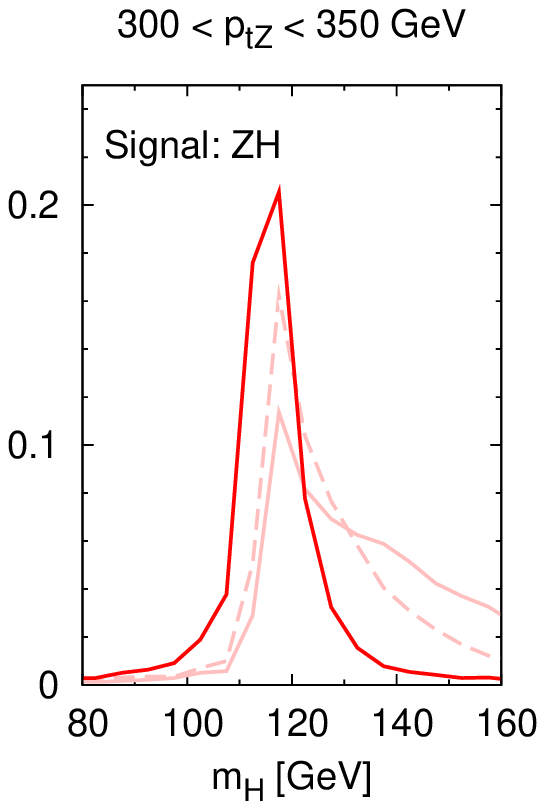}
  \includegraphics[width=0.45\linewidth]{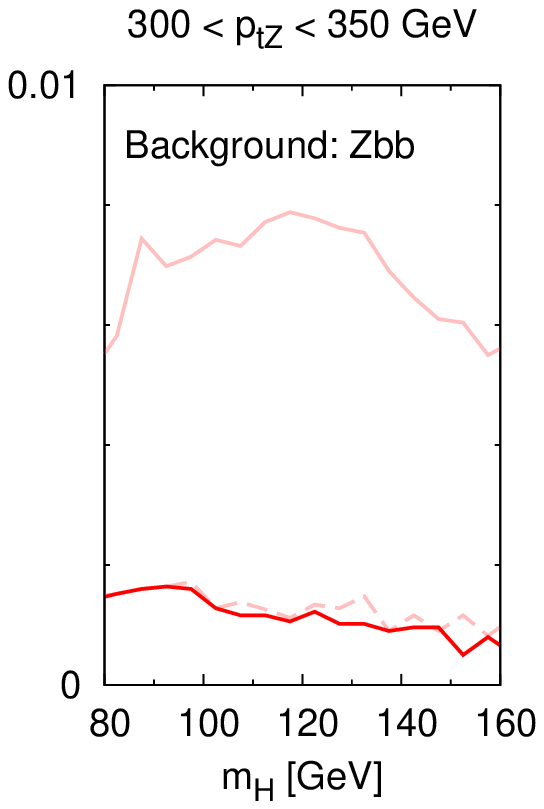}
  \caption{Invariant mass distribution of the hadronic part of
    high-$p_t$ ZH and $Zb\bar b$ events, for the hardest jet (light
    solid line, using $R=1.2$), after the mass drop (light dashed
    line) and after filtering (dark solid line). The normalisation is
    arbitrary (and unrelated) in the two plots.  Events simulated with
    Herwig~6.5~\cite{herwig2}, Jimmy~4.3~\cite{jimmy} and
    reconstructed with FastJet~2.3~\cite{Cacciari:2005hq}.}
  \label{fig:MDF-in-action}
\end{figure}

The procedure is summarised in fig.~\ref{fig:MDF}, and illustrative
invariant-mass distributions at the different stages are given in
fig.~\ref{fig:MDF-in-action}, showing how the mass-drop is the
critical stage for eliminating the background, while filtering is
crucial for obtaining good mass resolution on the signal.

\begin{figure}
  \centering
  \includegraphics[width=\linewidth]{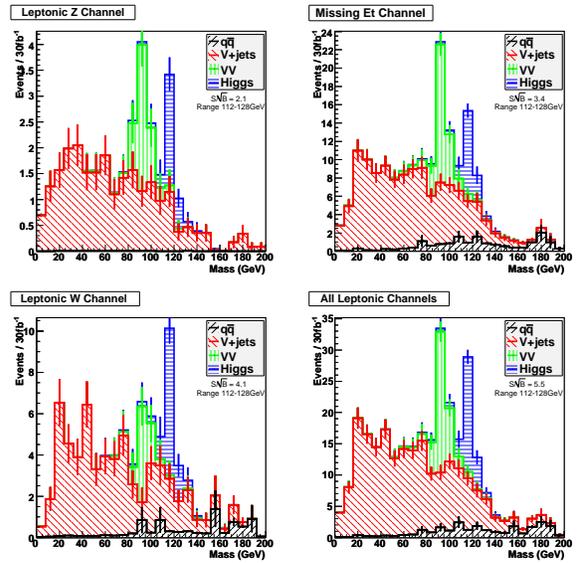}
  \caption{Simulated invariant mass distributions in three different
    channels, together with the combined results (bottom right).}
  \label{fig:300gev-res}
\end{figure}

\begin{figure}
  \centering
  \includegraphics[width=\linewidth]{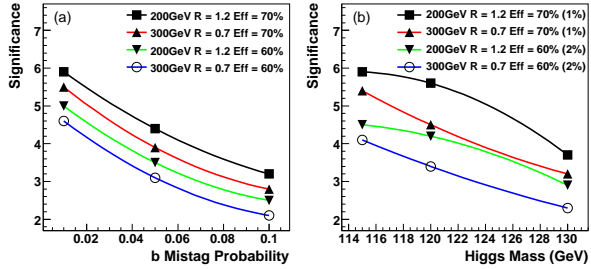}
  \caption{Dependence of the result for $S/\sqrt{B}$ on $b$ mistag
    rate (left) and on the Higgs-boson mass (right), for various
    combinations of $p_{t,min}$, $R$ and $b$-tagging efficiency.}
  \label{fig:curves}
\end{figure}

To test the potential of a high-$p_t$ VH analysis for Higgs discovery,
ref.~\cite{Butterworth:2008iy} considered simulated $VH$ events
($V=W,Z$), and backgrounds from $Vj$ (including $Vb \bar b$), $VV$,
$t\bar t$, single-top and dijet events, generated with
Herwig~\cite{herwig2} (and UE from Jimmy\cite{jimmy}).
%\footnote{Cross checks
%  were performed with MCFM (signal and $Vbb$)and MC@NLO (signal)
%  to verify that NLO $K$-factors leave the conclusions unchanged}
%
The precise results depend on the choice of $p_{t,min}$, $R$ for the
jet finder and the $b$-tagging efficiency and fake rate. The main
results of \cite{Butterworth:2008iy} were for $p_{t,min}=200\GeV$,
$R=1.2$ and $b_{\rm eff/fake} = 60\%/2\%$. Here we complement this with
results for $p_{t,min}=300\GeV$, $R=0.7$ and $b_{\rm eff/fake} =
70\%/1\%$, fig.~\ref{fig:300gev-res}, where the number of events
corresponds to $30\fb^{-1}$, without any $K$-factors.
This shows a clear mass peak around $m_H=115\GeV$, together with a
potentially very useful control peak from $VZ$ events with  $Z\to
b\bar b$. The value of ${\rm
  signal}/\sqrt{\rm background}$ is about $5.5$ for the combination of
all leptonic channels, based on a mass window of $16\GeV$, which is
roughly compatible with expected experimental resolutions.
The dependence of the result on the $b$-tagging scenario and the Higgs
mass is shown in fig.~\ref{fig:curves} and one sees that the channel
remains viable even with worse $b$-tagging and for masses up to $\sim
130\GeV$.

One may ask how it is that by throwing out $95\%$ of the signal one
can improve the chances of discovery compared to the analysis of
\cite{atlasphystdr}. The answer involves many aspects: for example the
$t\bar t$ background is nearly completely eliminated because it is 
difficult for $t\bar t$ events to produce a collimated $b\bar b$ pair
that recoils against a high-$p_t$ W-boson; other backgrounds fall
faster with $p_t$; signal acceptance and mass resolution improve at
high-$p_t$; and new signal channels arise ($Z \to \nu\nu$). The impact
of each effect is summarised in table~\ref{tab:gains-losses} (whose
entries are approximate because we have not fully repeated the
analysis of \cite{atlasphystdr}). Additionally, the high-$p_t$ analysis is
free of cut and top-induced artefacts, making it much easier to claim
discovery once one identifies a mass peak.

\begin{table}
  \begin{tabular}{lrr}\hline
          & Signal & Background \\ \hline
   Eliminate $t\bar t$, etc.  & $-$          & $\times 1/3$ \\
   $p_t > 200 \GeV$                 & $\times1/20$ & $\times
   1/60$\tablenote{for $Wb\bar b$ and $Zb\bar b$ backgrounds}\\
   improved acceptance              & $\times 4$   & $\times 4$     \\
   twice better resolution          & $-$          & $\times 1/2$   \\
   add $Z\to \nu\bar \nu$           & $\times 1.5$ & $\times 1.5$   \\
   \hline
    total                       & $\times 0.3$ &  $\times 0.017$
  \end{tabular}
  \caption{Approximate impact of different aspects of the high-$p_t$
    analysis on the signal and background in $W/Z$-associated Higgs
    production, relative to the low $p_t$ analysis.}
\label{tab:gains-losses}
\end{table}

In considering the above results it is important to bear in mind that
they are based on hadron-level simulation. Ultimately, this channel's
degree of competitiveness for Higgs discovery (notably compared to
$gg\to H \to\gamma\gamma$) will depend on the detailed detector
performance (studies are in progress), as well as possible
future improvements (e.g.\ separate consideration of $200 < p_{tV} <
300 \GeV$ and $p_{tV} > 300 \GeV$).
Nevertheless we believe that our choices at
hadron-level (e.g.\ the mass-window width) are sufficiently
conservative that there is a high likelihood that this will be
fruitful channel for LHC Higgs studies.
This is important as it is the only channel that can provide separate
measurements of $WH$ and $ZH$ couplings, and the control $Z$-peak will
provide a constraining standard candle for normalisation (especially
once all diboson production channels have been calculated to NNLO).

Finally, the analysis discussed here may have more general lessons:
one is that in searches dominated by large backgrounds with
cut-induced (or top-induced) artefacts in the mass distribution, going
to high $p_t$ can help limit their impact. Another is that other studies
involving highly-boosted heavy objects ($W,Z,H,t$, see
e.g.~\cite{AgasheSUSY08}) stand to benefit significantly from the new
mass-drop and filtering jet techniques developed here, as has already
been seen in a related high-$p_t$ $t\bar t$ analysis~\cite{Kaplan:2008ie}.

This work was supported in part by grant
ANR-05-JCJC-0046-01 from the French Agence Nationale de la
Recherche.

\bibliographystyle{aipproc}   % if natbib is available

\bibliography{eurostar}

\end{document}